\documentclass[runningheads]{llncs}
\usepackage[T1]{fontenc}

\usepackage{graphicx}
\usepackage{comment}
\usepackage{amsmath}
\usepackage{amssymb}
\usepackage{array}
\usepackage{enumitem}

\begin{document}

\title{Network-wide Quantum Key Distribution with Onion Routing Relay}


\author{Pedro Otero-García \and
David Pérez-Castro \and
Manuel Fernández-Veiga \and \\
Ana Fernández-Vilas}
\institute{atlanTTic - research center, Universidade de Vigo 
\email{\{pedro.otero,dperezcastro,mveiga,avilas\}@det.uvigo.es}}

\maketitle

\begin{abstract}
The advancement of quantum computing threatens classical cryptographic methods, necessitating the development of secure quantum key distribution (QKD) solutions for QKD Networks (QKDN). In this paper, a novel key distribution protocol, Onion Routing Relay (ORR), that integrates onion routing (OR) with post-quantum cryptography (PQC) in a key-relay (KR) model is evaluated for QKDNs. This approach increases the security by enhancing confidentiality, integrity, authenticity (CIA principles), and anonymity in quantum-secure communications. By employing PQC-based encapsulation, ORR aims to avoid the security risks posed by intermediate malicious nodes and ensures end-to-end security. Our results show a competitive performance of the basic ORR model, against current KR and trusted-node (TN) approaches, demonstrating its feasibility and applicability in high-security environments maintaining a consistent Quality of Service (QoS). The results also show that while basic ORR incurs higher encryption overhead, it provides substantial security improvements without significantly impacting the overall key distribution time. Nevertheless, the introduction of an end-to-end authentication extension (ORR-Ext) has a significant impact on the Quality of Service (QoS), thereby limiting its suitability to applications with stringent security requirements.
\end{abstract}

\section{Introduction \label{sec:intro}}

The development of quantum computing is rapidly evolving~\cite{aasen2025}, with recent advancements in quantum computation that brings algorithms~\cite{jordan2018} like Shor's and Grover's closer to reality. These advances pose a potential risk to existing cryptographic systems, as a quantum adversary could eventually break the security of classical cryptography, especially asymmetric algorithms.

The scientific community has focused on two main approaches to address quantum threats: quantum cryptography (QC), which uses quantum systems for security, and post-quantum cryptography (PQC), which seeks algorithms resilient to quantum attacks. QC leverages quantum mechanics to ensure security, with Quantum Key Distribution (QKD) being a prominent solution~\cite{bennett1984}, relying on quantum superposition or entanglement~\cite{ekert1991}. However, QKD faces implementation challenges~\cite{stanley2022}: The lack of commercial quantum repeaters or scalability constrains, both of which limit its range of applicability.

In contrast, PQC does not require specialized hardware and offers a more practical, cost-effective solution, although it is based on complex mathematical problems that could eventually be broken by sufficiently powerful quantum computers~\cite{alagic2025}. NIST is leading the effort to standardize PQC algorithms, including CRYSTALS-Kyber, CRYSTALS-Dilithium, HQC or FALCON.

Despite the hardware limitations of QC, QKD remains the only solution proven to be unconditionally secure against quantum attacks. However, scalability remains a challenge, with models like key-relay (KR)~\cite{elliott2002} and trusted-node (TN)~\cite{itu2020} not ensuring end-to-end confidentiality across intermediate nodes. Thus, a hybrid QC and PQC approach is required for comprehensive security in quantum key distribution networks (QKDNs).

This paper evaluates the feasibility of our previous work~\cite{otero2025} ---a secure key distribution approach based on the key-relay (KR) model, enhanced with Onion Routing (OR)~\cite{goldschlag1999} and PQC techniques to ensure CIA principles and anonymity--- by comparing it with the main existing alternatives: key-relay and trusted-node models. The remainder of the document is structured as follows: Section~\ref{sec:related} reviews existing QKDN security models, Section~\ref{sec:tech-bg} outlines the technical background, Section~\ref{sec:model} introduces the ORR model, Section~\ref{sec:tests} presents the experimental setup, and Section~\ref{sec:results} provides a comparative performance analysis. Finally, Section~\ref{sec:conclusions} summarizes the findings, highlighting ORR(-Ext)'s advantages and drawbacks in security and QoS ending with the future lines of work to follow.

\section{Related Work} \label{sec:related}

In QKDNs without commercial quantum repeaters, the primary approaches to overcome distance limitations are KR and TN models. Both require fully trusted nodes, introducing security vulnerabilities~\cite{huttner2022}. Rass \textit{et al.}~\cite{rass2024} emphasize that network security is only as strong as the least trusted intermediate node. De Santis \textit{et al.}~\cite{desantis2024} propose satellite-based configurations to reduce dependence on multiple intermediaries, assuming only one or two trusted nodes.
Calsi \textit{et al.}~\cite{licalsi2025} introduce a KR enhancement where nodes establish QKD links with both nearest and next-nearest neighbors, improving resilience but remaining susceptible to multiple-node attacks. In TN models, Vyas \textit{et al.}~\cite{vyas2024} suggest trust-level segmentation to relax security assumptions at intermediate nodes, though this requires additional infrastructure for trust verification.

To address malicious node threats, post-quantum cryptography (PQC) has been integrated into classical protocols~\cite{schwabe2020,shim2025}. Rios \textit{et al.}~\cite{rios2025} show that combining Kyber and Dilithium enhances classical crypto performance at high-security levels, albeit with increased traffic overhead.
Hybrid QC–PQC solutions are increasingly favored for critical infrastructure in the NISQ era~\cite{zeng2024}, as they offer both long- and short-term protection~\cite{wang2021}. One approach uses QKD to encrypt PQC-based key exchanges~\cite{djordjevic2020}.

Building on this trend, we adopt a hybrid method combining KR-based QKD with OR and PQC, enabling secure key distribution between distant nodes, as introduced in our earlier work~\cite{otero2025}.

\section{Technological Background} \label{sec:tech-bg}

QKD and PQC derive security from fundamentally different principles. QKD offers unconditional security based on quantum mechanics, while PQC achieves computational security against quantum attacks. While QKD ensures point-to-point confidentiality, PQC excels in speed and scalability. By combining both, one can build resilient systems for both short- and long-term security. Our method incorporates OR and PQC-encrypted keys into a KR-based QKDN, achieving layered protection and preserving end-to-end confidentiality.

\subsection{Key-Relay Quantum Key Distribution Networks}

In KR networks, a sender and receiver share a secret key $S$ via a path of intermediate QKD-enabled nodes. The sender generates $S$ using a QRNG, encrypts it with the shared quantum key, and transmits it through the network. Each node decrypts and re-encrypts $S$ for the next hop. Although simple and effective, this model exposes $S$ to every intermediate node, compromising confidentiality.
KR is also vulnerable to spoofing on classical channels. Although intercepting a quantum key is unlikely without compromising a QKD node, such attacks can still disrupt network operations.

\subsection{Onion Routing}\label{sec:onion}

OR is a layered encryption technique designed to anonymize communication paths. A message is encrypted in successive layers, starting with the public key of the final node and moving backward to the first one. Each node decrypts one layer, learns only the next hop, and forwards the message, preserving sender-receiver anonymity.
OR enhances privacy but introduces higher latency and exposes the exit node to potential attacks. Some extensions~\cite{kuhn2019} improve security by verifying message integrity end-to-end, defending against message tampering and overload-based denial-of-service (DoS) attacks on onion routers, however, increasing the computation and the size of the message.

\section{Model Principles} \label{sec:model}

\begin{figure}
    \centering
    \includegraphics[width=\linewidth]{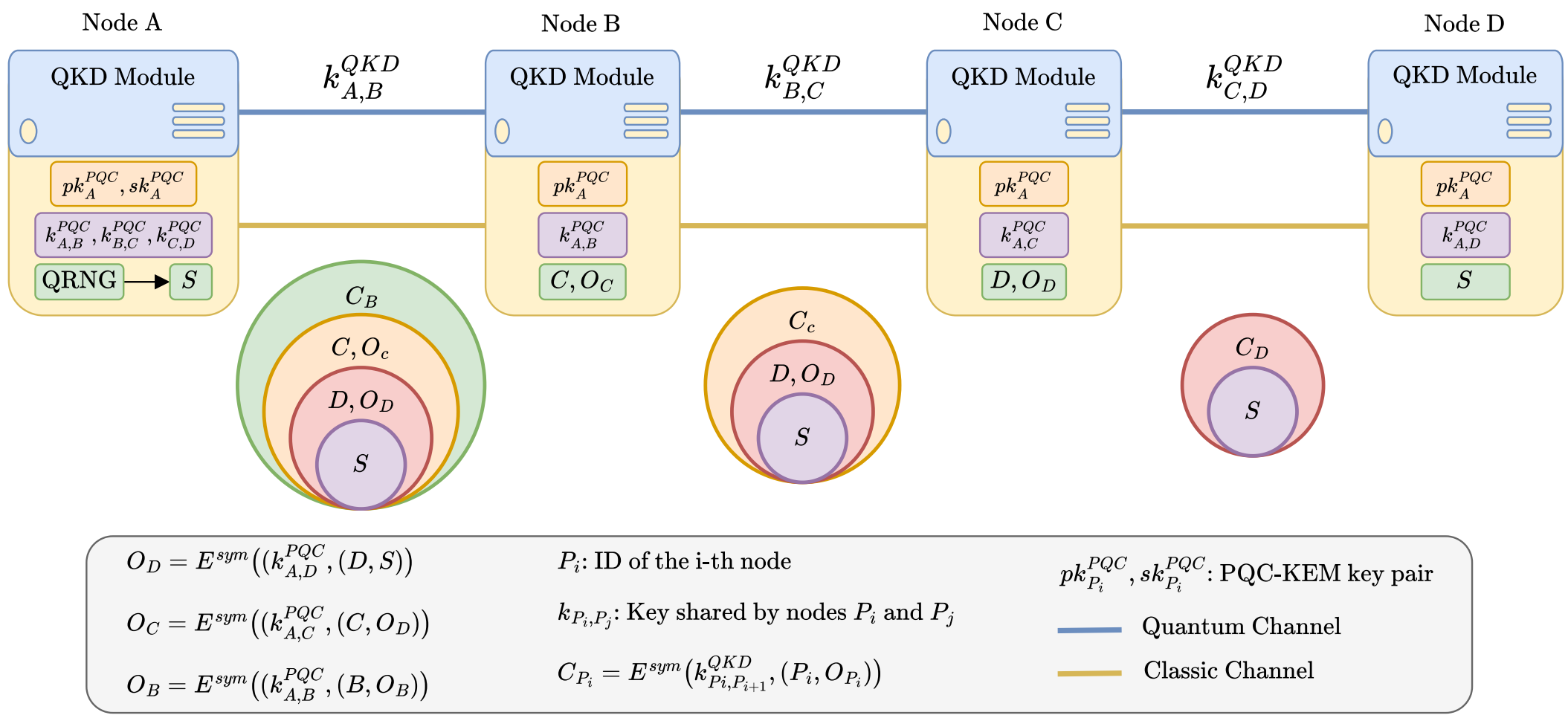}
    \caption{Simplified example of the Onion Routing Relay key distribution model}
    \label{fig:ORR}
\end{figure}

The proposed model, Onion Routing Relay (ORR), guarantees unconditionally secure key distribution within QKDN. It builds on the KR model for QKDN and integrates an OR extension that ensures end-to-end authentication. This approach upholds CIA principles and anonymity of the destination node. Figure~\ref{fig:ORR} illustrates the basic operation of ORR.

Since classical OR protocols remain vulnerable to quantum attacks, the model replaces insecure classical cryptography ---particularly asymmetric cryptography--- with PQC. Rather than using public-key cryptography to encrypt the onion layers, ORR uses PQC Key Encapsulation Mechanism (PQC-KEM) algorithms to distribute keys among nodes. It then applies symmetric encryption algorithms, such as AES,\footnote{Due to Grover's algorithm halving the security of symmetric cryptography, AES-256 should be used in a post-quantum world to maintain the same security level as AES-128 today.} to create the encrypted layers (onions). The ORR model is detailed described in our previous work~\cite{otero2025}.

\section{Test scenario \& Implementation} \label{sec:tests}

To compare the different models, a series of scripts were developed in C to simulate their behavior. A GitHub repository\footnote{\url{https://github.com/pedrotega/ORRvsTNvsKR}} is available to readers, providing the source code and execution instructions. It is important to note that due to the performance differences between the ORR model and its extended version with end-to-end authentication (ORR-Ext), it has been decided to evaluate each approach separately.

For the experiment, only a pair of QKD nodes was available --specifically the Cerberis QKD XGR~\cite{idq2021} developed by ID Quantique-- which provided the quantum functionalities for the simulations. Each simulated node in the scripts requests QKD keys from one of the real nodes, which works as a key-managment-system, and retrieves the key by its identifier from the other real node. All the request follows a RESTful API format specified in ETSI-014~\cite{etsi2019}.

In the code, each node is executed in a different thread that allows  the execution of code concurrently or sequentially, as appropriate, at any given time. Since a QRNG was not available for this project, a PRNG (Pseudo-Random Number Generator) was used as a replacement. Table~\ref{tab:crypto_algorithms} lists the algorithms and functions employed in the implementation, along with the respective software suppliers.

\begin{table}[t]
    \centering
    \caption{Cryptographic Algorithms and Their Implementations}
    \begin{tabular}{|c|c|c|c|}
        \hline
        \textbf{Category} & \textbf{Implementation} & \textbf{Category} & \textbf{Implementation} \\ 
        \hline
        Symmetric Encryption & AES-256-CBC (OpenSSL) & PQC-KEM & Kyber-768 (LibOQS) \\
        \hline
        PRNG & RAND\_bytes() (OpenSSL) & XOR & Directly in C \\ 
        \hline
    \end{tabular}
    \label{tab:crypto_algorithms}
\end{table}

The parameters measured in the tests include the encryption time and the key distribution time, which is calculated from the moment the secret is generated at the initiator node until it reaches the destination node. For the encryption time, the measurement varies depending on the model: In ORR the longest encryption time in this is spent encrypting the initial onion at the initiator node. In the version with authentication end-to-end (ORR-Ext) the time to create the authentication extension is included. For the TN model, the encryption time is measured during the calculation of the ciphertext to be sent to the destination node at the trusted node. Finally, in the KR model it is only necessary to measure the time of XOR operation when encrypting the secret in the initiator node.

For the implementation of ORR-Ext, the HMAC-256 algorithm was employed as signature scheme. This choice is motivated by the fact that it is still considered secure even in the presence of Grover's algorithm, offering post-quantum resistance without the overhead associated with PQC alternatives. In contrast to HMAC, PQC signature schemes are asymmetric --which necessitates the inclusion of the public key in each message-- significantly increasing the overall message size. Table~\ref{tab:onion-len} presents a comparison of the message lengths in both ORR and ORR-Ext models under different signature algorithms, highlighting the trade-offs in communication efficiency introduced by PQC signature methods.

\begin{table}[t]
    \centering
    \caption{Ciphertext length (Bytes) for a circuit of 5 nodes}
    \begin{tabular}{|c|c|c|c|c|}
        \hline
        \textbf{Variant} & \textbf{Onion} & \textbf{Public Key} & \textbf{Signature} & \textbf{Ciphertext}\\ 
        \hline
        ORR & 416 & 0 & 0 &  432\\
        \hline
        ORR-Ext-HMAC-256 & 416 & 0 & 32 &  1472\\
        \hline
        ORR-Ext-Falcon-1024 & 416 & 1793 & 1280 &  16612\\
        \hline
        ORR-Ext-Dilithium3 & 416 & 1952 & 3293 &  27537\\
        \hline
    \end{tabular}
    \label{tab:onion-len}
\end{table}


\section{Results and Discussion} \label{sec:results}

\begin{figure}[t]
    \centering
    \includegraphics[width=0.495\linewidth]{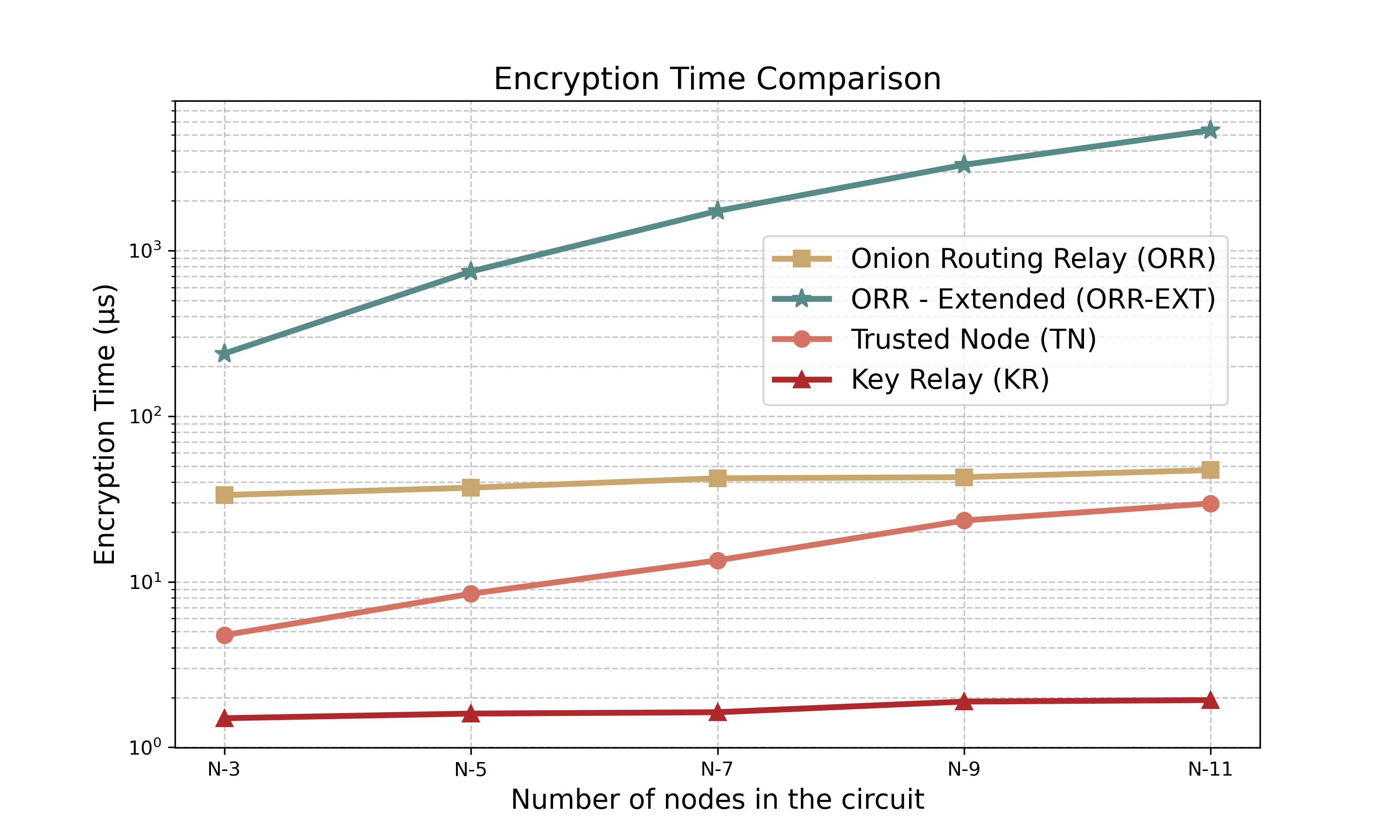}
    \includegraphics[width=0.495\linewidth]{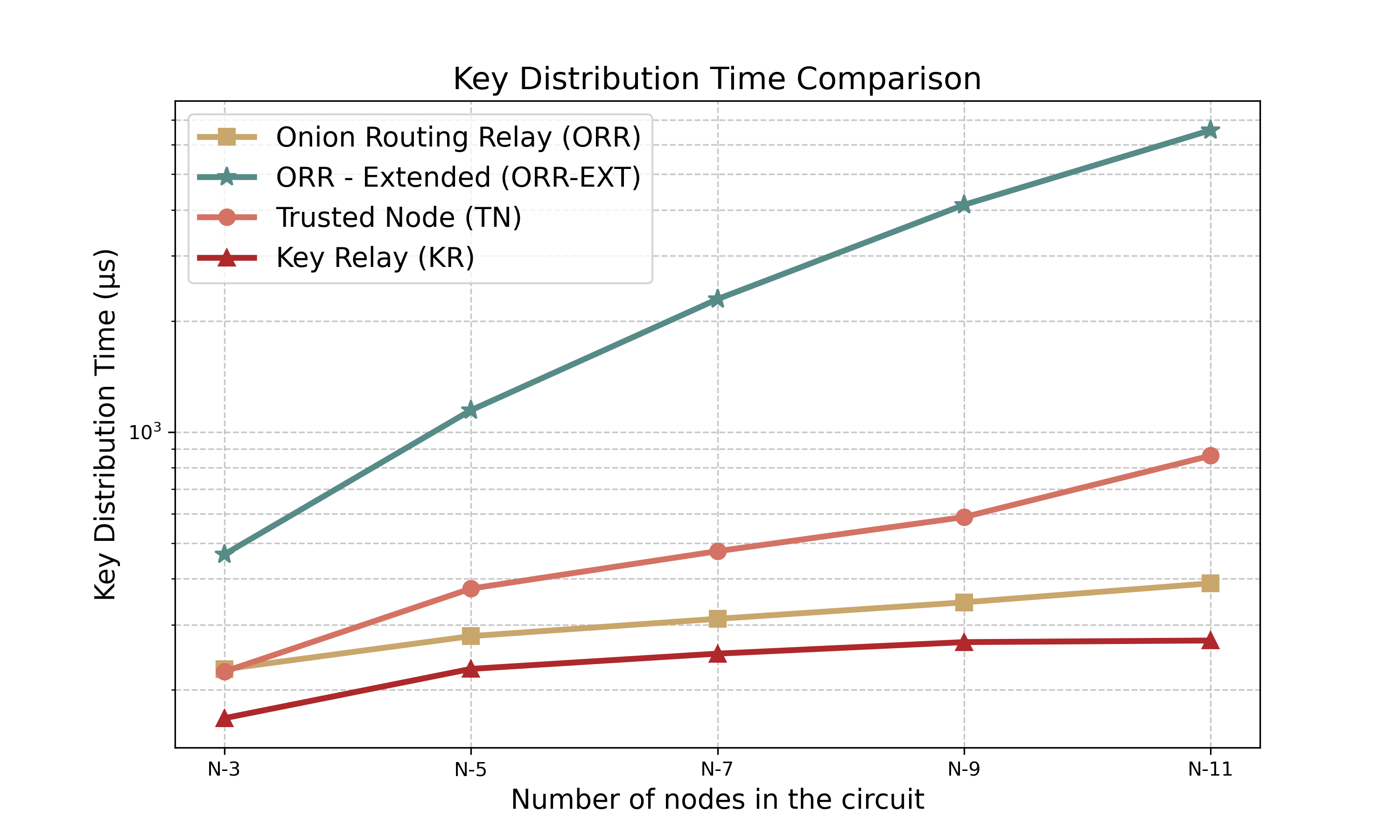}
    \caption{Encryption time \& key distribution time comparison of the different models}
    \label{fig:comparison}
\end{figure}

The results presented in this section have been obtained by averaging 100 iterations with each simulated model for circuits\footnote{A circuit is formed by the intermediate and destination nodes.} with 3, 5, 7, 9, and 11 nodes. Figure~\ref{fig:comparison} (left) illustrates the average time required for the encryption procedure in each model. The KR model exhibits a highly consistent and minimal encryption time, ranging from 1.5 to 1.93~$\mu s$, remaining largely unaffected by the number of nodes due to its simple XOR-based encryption using QKD keys. The TN model shows a moderate increase, starting at 4.76~$\mu s$ for a 3-node circuit and reaching 29.7~$\mu s$ at 11 nodes, as encryption requires handling more ciphertexts with each additional node. The ORR model, which implements layered encryption, maintains moderate growth in encryption time, starting at 33.4~$\mu s$ and increasing up to 47.25~$\mu s$ with 11 nodes. Meanwhile, the ORR-Ext model introduces a significant overhead due to the additional operations required to ensure end-to-end authentication. It starts at 238.78~$\mu s$ for 3 nodes and escalates sharply to 5327.31~$\mu s$ for 11 nodes.

Regarding the key distribution time, represented in Figure~\ref{fig:comparison} (right), the growth trend across models reflects the increasing communication and coordination overhead as the number of nodes rises. The KR model again stands out as the fastest, beginning at 167.7~$\mu s$ for 3 nodes and reaching 272.31~$\mu s$ for the largest circuit. The TN model, which starts slightly above KR at 224.24~$\mu s$, experiences more dramatic growth, requiring 862.55~$\mu s$ at 11 nodes. The ORR model, while initially slower than KR and TN (227.62~$\mu s$), maintains a steadier increase, finishing at 389.02~$\mu s$. In contrast, the ORR-Ext model incurs the highest key distribution times due to the authentication processes involved and the major length of their ciphertexts, with the average time rising from 465.8~$\mu s$ for 3 nodes to 6571.89~$\mu s$ for the 11-node configuration.

From the results obtained in the previous section, it can be concluded that the inclusion of layered encryption and authentication mechanisms results in significantly different performance profiles among the four models. As expected, the ORR-Ext model introduces the most substantial encryption overhead, requiring several orders of magnitude more time compared to the TN and KR models. Specifically, the encryption time in ORR-Ext can be more than 1800 times higher than in KR and over 150 times higher than in TN for the 11-node circuit. This is due to the added authentication extension operations needed to guarantee end-to-end security, which accumulate across layers.

The TN model shows a moderate but clear growth in encryption time as the number of nodes increases, stemming from the need to aggregate and process ciphertexts at the trusted node. This processing overhead is absent in the KR model, which maintains consistently minimal encryption times, as its XOR operation does not depend on the circuit size. The basic ORR model without the authentication extension, while implementing layered encryption, shows a more controlled increase, demonstrating its balance between added security and computational efficiency compared to ORR-Ext.

Nevertheless, the most impactful distinction across models lies in the key distribution time, which directly affects the system's QoS. The ORR-Ext model, in particular, suffers from very high distribution times due to both the authentication-related message exchanges and the context switching between threads simulating node-to-node communication. These results suggest that in practical deployments, where network delays and message propagation times are even more significant, the additional burden introduced by ORR-Ext could represent a limiting factor in terms of responsiveness. TN also exhibits a steep rise in key distribution time, particularly for larger circuits, primarily due to the additional communication step required to forward the key to the final destination.

In contrast, the KR and basic ORR models display a more favorable trend, especially at higher node counts. Notably, while basic ORR's encryption time is higher than that of KR, its key distribution time remains relatively competitive. For example, in the 11-node case, the difference between ORR and KR in key distribution is approximately 117~$\mu s$, which is consistent with the expected communication overhead introduced by layered routing, yet still acceptable within practical bounds.

Therefore, it can be concluded that although ORR-Ext significantly enhances security guarantees through authentication, it does so at the cost of scalability and performance. In realistic environments where encryption times become negligible compared to network latency, models like ORR—-despite their layered encryption—-can achieve a competitive QoS while offering enhanced protection against malicious intermediaries, presenting a viable compromise between efficiency and security.

\section{Conclusions and Future work} \label{sec:conclusions}

In this paper, we have proposed and tested a key relaying solution based on the principles of onion routing, to ensure the anonymity of end-to-end keys and the sequence of intermediate nodes. Our evaluation shows that the cost of nested PQC encryption with an ORR scheme is high but affordable if the extra level of security in relaying must be achieved. Therefore, ORR and ORR-Ext both strengthen the security of key relaying in QKD networks at the cost of some extra bandwidth only. Future work will focus on implementing these models in a real QKDN to evaluate whether the proposed solutions can meet the requirements of practical deployment scenarios. In particular, the objective is to determine whether ORR-Ext can sustain a minimum key distribution rate that would make it suitable for highly security-sensitive applications. \\

\noindent \textbf{Acknowledgments} This publication is part of the I+D+I project "Creation of a network of laboratories and demonstration centers for research and innovation in cybersecurity solutions." financed, by “European Union NextGeneration, the Recovery Plan, transformation and Resilience, and INCIBE”.

\end{document}